\documentstyle[12pt,aps,epsfig]{revtex}

\newcommand{\be}{\begin{equation}}
\newcommand{\ee}{\end{equation}}
\newcommand{\bea}{\begin{eqnarray}}
\newcommand{\eea}{\end{eqnarray}}
\newcommand{\pv}{{\bf p}}
\newcommand{\kv}{{\bf k}}

\newcommand{\pvuni}{\hat{{\bf p}}}

\newcommand{\puni}{\hat{p}}

\newcommand{\kuni}{\hat{k}}

\newcommand{\cO}{{\cal O}}

\newcommand{\GG}{{\tilde G}}

\newcommand{\Tr}{\mbox{Tr}}

\def\siml{\; \raise0.3ex\hbox{$<$\kern-0.75em
      \raise-1.1ex\hbox{$\sim$}}\; }
\def\simg{\; \raise0.3ex\hbox{$>$\kern-0.75em
      \raise-1.1ex\hbox{$\sim$}}\; }

\begin{document}

\preprint{EHU-FT/000?}
\draft
\title{Matter-induced vertices for photon splitting \\ in a weakly magnetized plasma}
\author{J. M. Mart\'\i nez Resco \thanks{\tt wtbmarej@lg.ehu.es}, \\ 
        M. A. Valle Basagoiti \thanks{\tt wtpvabam@lg.ehu.es}}

\address{Departamento de F\'\i sica Te\'orica, \\ 
 Universidad del Pa\'\i s Vasco, Apartado 644, E-48080 Bilbao, Spain}

\date{\today}

\maketitle

\begin{abstract}

We evaluate the three-photon vertex functions at order $B$ and $B^{2}$ in a weak 
constant magnetic field at finite temperature and density with on shell external 
lines. Their application to the study of the photon splitting process leads to 
consider high energy photons whose dispersion relations are not changed significantly 
by the plasma effects. The absorption coefficient is computed and compared with 
the perturbative vacuum result. For the values of temperature and density of 
some astrophysical objects with a weak magnetic field, the matter effects are 
negligible. 

\end{abstract}   



\newpage
\section{Introduction}

The photon splitting process in the presence of an external magnetic field was 
thoroughly studied in the vacuum by Adler~\cite{adler}. Since then, it has been 
studied several times over the last years using different techniques and for 
different regimes~\cite{baier,adler1,baring,chistyakov} due to its possible 
relevance in astrophysical applications (see e.g.\cite{harding}). 

In the vacuum, the Furry theorem forbids the processes with an odd number of 
photon vertices. The lowest order contribution comes from the box graph with an 
insertion of the external magnetic field. In the collinear aproximation (which 
is always considered in the rest of this paper), this matrix element vanishes 
due to Lorentz and gauge invariance. Thus, the first non-zero contribution to 
the amplitude comes from the hexagon which is of order $B^{3}$. Adler also studied 
how the propagation of photons is modified by the presence of a magnetic field. 
He found that there are two linearly polarized modes and due to CP invariance, 
all processes with an odd number of perpendicularly polarized photons vanish. 
This along with the induced  birefringence in the polarized vacuum enforces that 
the only allowed channel is $\parallel \to \perp \perp$. The absorption coefficient 
is 
\be
 \kappa=\frac{\alpha^{3}}{15 \pi^{2}}\left(\frac{13}{315}\right)^{2}
 \left(\frac{\omega}{m}\right)^{5}
 \left(\frac{B \sin \theta}{B_{cr}}\right)^{6}m,
\ee
where $\theta$ is the angle between the magnetic field and the direction of the 
incident photon, $\omega$ is the energy of the incident photon and 
$\vert e \vert B_{cr}=m^{2} \simeq 4.4 \:  10^{9} \ T$. 

In the presence of a hot or dense medium, explicit Lorentz invariance is lost 
since now there is a privileged reference frame, namely the frame in which the 
system is at rest. Therefore the box diagram does not need to vanish. In addition, 
for a non zero net electron density, the ground state is not C invariant and 
hence the amplitudes with an odd number of photon vertices do not need to vanish 
either. Since the environment in astrophysical objects is not exactly the vacuum, 
it is worth studying the effects that density and temperature can have in the 
photon splitting process.

As far as we know, a quantitative analysis of the matter effects on the photon 
splitting has been reported in a few papers~\cite{elmfors,gies,bulik}. In 
Ref.~\cite{bulik}, the absorption coefficient is computed by including the plasma 
effects on the dispersion relations but using the vacuum vertices from the 
Euler-Heisenberg effective Lagrangian. In Ref.~\cite{elmfors}, a one-loop thermal 
effective Lagrangian for a constant electromagnetic field is computed and from 
that the relevant matrix element of order $B$ is read by following the same 
procedure that is used in the absence of a medium~\cite{adler,landau}. In 
Ref.~\cite{gies}, the same approach is used starting from a two-loop effective 
Lagrangian. Thus, the amplitudes reported in Refs.~\cite{elmfors,gies} have an 
analytic dependence on the frequencies of the photons in the process. While this 
approach works properly in the vacuum, it can have a flaw when it is applied for 
the study of the plasma effects. 

In the vacuum, a consequence of Lorentz and gauge invariance is that the effective 
action for the electromagnetic field displays an analytic dependence on the field 
strength and its derivatives. It is just this analytic behaviour which enables a 
well defined derivative expansion of the effective action whose first few terms 
of the Taylor's series can be computed from the Euler-Heisenberg Lagrangian
\footnote{When the masses are different from zero and $n \geq 3$, the one-particle 
irreducible $n$-point functions are analytic functions of all its arguments, 
$p_{i}^2 = \omega_{i}^2 - \pv_{i}^2$ and 
$p_{i} \cdot p_{j} = \omega_{i} \omega_{j} -\pv_{i}\cdot \pv_{j}$ for  
$(p_1,\ldots, p_{n-1}) = (0,\ldots, 0)$, so that its Taylor series can be guessed 
from the behaviour for $\omega_{i}=0$.}. Moreover, because of  Adler's theorem, 
the three-point photon vertex in the collinear approximation computed from the 
Euler-Heisenberg Lagrangian is exact to order $B^3$.

On the other hand, in a medium, Lorentz invariance is lost and the effective 
action displays a non-analytic behaviour~\cite{weldon,das}. This invalidates 
the identification of proper photon vertices based on an effective Lagrangian 
computed by a derivative expansion technique. In order to clarify this essential 
point, let us consider the thermal two-point polarization tensor in QED at high 
temperature. In the static limit, the only non-zero component of this tensor is 
$\Pi^{00}(\omega=0, \kv)=-m_{D}^{2}+\cO(k^2)$, where $m_{D}^{2}=e^2 T^2/3$ is the 
Debye mass of static screening. Then, the corresponding lowest order terms of a 
derivative expansion in the effective action are 
$\Gamma \sim \int d^4 x \left(\frac{1}{2} m_{D}^2 A_{0}^2
+\cO({\mathbf E}^2)+\cO({\mathbf B}^2)\right)$. Here $A_{0},{\bf E}$ and ${\bf B}$
represent the temporal component of the gauge field, the electric and the 
magnetic field respectively.
However, when $\omega \approx k$, the non-zero components are 
$\Pi^{ij}(\omega=k, \kv)=m_{t}^2 (\delta^{i j} - \kuni^i \kuni^j)$, where 
$m_{t}^2=e^2 T^2/6$ is the transverse photon mass defined below in Eq.~(\ref{mt}). 
Now, the corresponding effective action is 
\be   \label{nlocal}
  \Gamma \sim \int d^4 x\, \frac{m_{t}^2}{2} \left( {\mathbf A}^2 + 
  {\mathrm div}{\mathbf A}\, \frac{1}{\nabla^2}\, {\mathrm div}{\mathbf A} \right), 
\ee
which has a non-local dependence in the fields but is gauge invariant. This 
differs from the corresponding effective action based on the derivative expansion 
and would not have been obtained from an expansion about a constant field. 
Certainly, in order to describe a dynamical process involving propagation of 
photons with a dispersion relation $\omega^2 \approx k^2$, the relevant matrix 
elements should be computed from an  effective action similar to the one in the 
Eq.~(\ref{nlocal}).
 
In this paper, we will directly compute the three point amplitude using the 
Schwinger propagator for an electron in an external magnetic field when the 
collinear external momenta are on shell. Expanding the three-photon vertex at 
lowest order in the external field and retaining the relevant pieces we will 
obtain the box contribution. Proceeding in the same way, we will also obtain the 
order $B^{2}$ contribution which would come from the pentagon graph. This leads 
to Eqs.~(\ref{triangleB}) and~(\ref{triangleB2}) below, which are the main 
findings of this paper. Both of them display a non-analytic dependence on the 
photon energies. By computing several components of the amplitudes, we have 
verified that the Ward identities are satisfied.  

The presence of a medium not only affects the amplitudes, allowing some of them 
not to vanish, but also modifies the photon propagation properties. However, at 
high energy we expect this modification to be negligible compared with that due 
to the magnetic field. In order to estimate under which conditions this is true, 
we recall that the dispersion relation of high energy photons in the absence of 
an external field is $\omega \simeq \sqrt{p^{2}+m_t^2}$, where $m_{t}$ is known 
as the transverse mass of the photon~\cite{braaten}
\be  \label{mt}
 m_{t}^{2}=\frac{4 \alpha}{\pi}\int_{0}^{\infty}\! \! dp \,
 \frac{p^{2}}{\varepsilon_{p}}\,
 (n_{F}(\varepsilon_{p}-\mu)+n_{F}(\varepsilon_{p}+\mu)), 
\ee
with $n_{F}(x)=1/(e^{\beta x}+1)$ and $\varepsilon_{p}=\sqrt{p^{2}+m^{2}}$ the 
energy of an electron of momentum $p$. The transverse mass is of the same order
as the plasma frequency. The refractive index is found to be 
$n \approx 1 - \frac{m_t^2}{2 p^2}$. On the other hand, the refractive indices 
due to vacuum birefringence are~\cite{adler}
\be  \label{refraction}
 n_{\parallel}^{\perp}=1+ \alpha a_{\parallel}^{\perp}
 \left(\frac{B \sin \theta}{B_{cr}}\right)^{2},\ \ \ 
 a_{\parallel}=\frac{4}{90\pi},\ \     a^{\perp}=\frac{7}{90\pi} .
\ee
In this paper we follow the convention of Adler~\cite{adler}, where $\perp$ ($\parallel$) 
means the magnetic field of the photon is perpendicular (parallel) to the plane
containing the external magnetic field and the photon's momentum vector.
Thus, we must require
\be\label{masst}
 p \simg \frac{1}{\sqrt{2\alpha a}}
 \left(\frac{B_{cr}}{B\sin \theta}\right) m_{t} \ \ \longrightarrow  \ \ 
 p \gg 10^{2} \left(\frac{B_{cr}}{B}\right) m_{t} \equiv \omega_{0} . 
\ee
This inequality specifies the condition that is necessary in order to neglect 
the medium effects on the photon propagation. We will find that in these conditions 
the contribution of plasma effects to the absorption coefficient is smaller than 
the vacuum contribution.  
 
In the next section we describe the computation of the box and pentagon amplitudes. 
The following section is devoted to obtaining the absorption coefficient and 
evaluating it in several situations. Finally in section IV we give a brief summary.


\section{The box and pentagon amplitudes}

In order to compute the three photon amplitude we need the fermion propagator 
in the presence of an external magnetic field. It was computed by Schwinger a 
long time ago~\cite{schwinger} using the proper time method 
\bea    \label{schwinger}  
 \! \! G(x,x')&=& 
 \! -i(i\! \! \not\!\partial-e\not\!\!A+m)\int_{-\infty}^{0}\! \! \! ds \, 
 \frac{-i}{(4 \pi)^{2}s^{2}}\: e^{i(m^{2}-i\epsilon)s} 
 e^{\Phi (x,x')} \: e^{i\frac{e}{2} \sigma^{\mu\nu}F_{\mu\nu}s}   \\  \nonumber
 && \times \exp \left( -\frac{1}{2} \Tr \ln[(eFs)^{-1} \sinh (eFs)]
 +\frac{i}{4}(x-x')eF\coth(eFs)(x-x') \right),  
\eea
with
\be
 \Phi (x,x')\! \equiv -i e \int_{x'}^{x} d \tilde{x}^{\mu} 
 (A_{\mu}(\tilde{x})+\frac{1}{2}F_{\mu\nu}(\tilde{x}-x')^{\nu}).
\ee
It can be expressed as 
\be
 G(x,x')=e^{\Phi (x,x')}\; \GG (x-x'),
\ee
where $\GG (x-x')$ is explicitly invariant under translations while the phase 
factor is not. We need the expansion of $\GG$ in powers of the magnetic field, 
for which we obtain in Fourier space
\bea  
 &&\GG(p)=\GG_{0}(p)+\GG_{1}(p)+\GG_{2}(p)+\cO(B^{3}),  \\  \nonumber
 &&\GG_{0}(p)=\frac{m {\bf 1}+\not \! p}{p^{2}-m^{2}},   \\  \nonumber
 &&\GG_{1}(p)=\frac{e}{(p^{2}-m^{2})^{2}} ({\tilde F}_{\mu\nu}\gamma^{\mu}\gamma^{5} p^{\nu}
 +\frac{m}{2} F_{\mu\nu}\sigma^{\mu\nu}),  \\  \nonumber
 &&\GG_{2}(p)=-2 e^{2}\frac{m {\bf 1}+\not \! p}{(p^{2}-m^{2})^{4}}
 (F_{\mu\alpha}F^{\alpha}_{\ \nu} p^{\mu}p^{\nu})
 +\frac{2 e^{2}}{(p^{2}-m^{2})^{3}}(F_{\mu\alpha}F^{\alpha}_{\ \nu}\gamma^{\mu}p^{\nu}),
\eea
with ${\tilde F}_{\mu\nu}=\varepsilon_{\mu\nu\alpha\beta}F^{\alpha\beta}/2$ and 
$\sigma^{\mu\nu}=\frac{i}{2}[\gamma^{\mu},\gamma^{\nu}]$. We have arrived at 
these results by expanding the integrand of $\GG$ in powers of $B$, performing 
the Fourier transform and finally doing the integral over the proper-time 
parameter. The expansion of $\GG$ in powers of $B$ had been carried out previously 
in Ref.~\cite{chyi}, where it was pointed out that the phase factor gives a non 
trivial contribution to the three-point amplitude
\bea
 e^{\Phi (y,x)} \cdot e^{\Phi (x,z)} \cdot e^{\Phi (z,y)}=
 \exp \left(\frac{ie}{2}(z-y)^{\alpha}F_{\alpha\beta}(y-x)^{\beta}\right)
 \\
 =e^{\Phi (x,z)} \cdot e^{\Phi (z,y)} \cdot e^{\Phi (y,x)}=
 \exp \left(\frac{ie}{2}(y-x)^{\alpha}F_{\alpha\beta}(x-z)^{\beta}\right)
 \nonumber \\  \nonumber
 =e^{\Phi (z,y)} \cdot e^{\Phi (y,x)} \cdot e^{\Phi (x,z)}=
 \exp \left(\frac{ie}{2}(x-z)^{\alpha}F_{\alpha\beta}(z-y)^{\beta}\right).
\eea 
Any of the three forms can be used in the computation of the amplitude due to 
the trace over the Dirac indices. Since the phase is already linear in $B$, there 
are two kinds of contributions at first order: one when there are two factors 
$\GG_{0}$ and one $\GG_{1}$ and the other when there are three factors $\GG_{0}$ 
and the phase. 

We start with the contribution without the phase which can be written directly 
in Fourier space (see Fig. 1)
\bea    \label{triangle}
 M_{1}^{\rho\mu\nu}[p,p_{1},p_{2}]&=& \! -(i)^{3} e^{3} T \sum_{n}\int
 \frac{d^{3}k}{(2 \pi)^{3}} \Tr[ \gamma^{\mu}\GG(k+p_{1}) \gamma^{\rho} \GG(k-p_{2})
 \gamma^{\nu} \GG(k)]  \\  \nonumber
 &&+(\, \mu,p_{1} \leftrightarrow \nu,p_{2}).  
\eea
To proceed we choose the momentum of the incident photon along the $OZ$ axis: 
$p^{\mu}=(\omega,0,0,p=\omega)$ and the magnetic field in the plane $OYZ$: 
${\bf B}=(0,B_{y},B_{z})$ without loss of generality. Thus, in the collinear 
aproximation the momentum of the outgoing photons are 
$p_{1}^{\mu}=(\omega_{1},0,0,p_{1}=\omega_1)$ and 
$p_{2}^{\mu}=(\omega_{2},0,0,p_{2}=\omega_2)$. 
The polarization states are chosen following the convention of Adler~\cite{adler}: 
$\epsilon_{\parallel}^{\mu}=(0,1,0,0)$, $\epsilon_{\perp}^{\mu}=(0,0,1,0)$. 

Introducing the expression for $\GG(p)$ in Eq.~(\ref{triangle}) and taking the 
linear part in $B$ we can compute the amplitude for the different channels. We 
have written a Mathematica code to carry out the Matsubara sums and the angular 
integrals. The Matsubara sums lead to derivatives of the Fermi distribution 
function which are integrated by parts, resulting at the end in the single 
integral of Eq.~(\ref{mt}) defining the transverse mass. We obtain 
\bea
 && M_{1}^{\parallel \to \perp \perp}=
 -\frac{1}{2} \frac{e B_{y}}{m^{2}}\frac{e}{\omega} \: m_{t}^{2},
 \\
 && M_{1}^{\perp \to \parallel \, \perp}=
 \frac{1}{2} \frac{e B_{y}}{m^{2}}\frac{e}{\omega_{1}} \: m_{t}^{2},
 \\
 && M_{1}^{\perp \to \perp \parallel}=
 \frac{1}{2} \frac{e B_{y}}{m^{2}}\frac{e}{\omega_{2}} \: m_{t}^{2},
 \\
 && M_{1}^{\parallel \to \parallel \, \parallel}=
 \frac{1}{2} \frac{e B_{y}}{m^{2}}\frac{e(\omega_{1}^{2}+\omega_{1}
 \omega_{2}+\omega_{2}^{2})}{\omega \omega_{1} \omega_{2}} \: m_{t}^{2},
 \\
 && M_{1}^{\parallel \to \parallel \, \perp}=M_{1}^{\parallel \to \perp \parallel}=
 M_{1}^{\perp \to \parallel \, \parallel}=M_{1}^{\perp \to \perp \perp}=0.
\eea

Now, in order to compute the contribution of the phase, we write the three-point 
function with the external legs in position space
\bea   \label{phasecont}
 \Gamma^{\rho\mu\nu}[x,y,z]= -(i)^{3} e^{3} 
 \int d^{4}x'd^{4}y'd^{4}z' 
 \Tr[
 \gamma^{\mu '}\GG_{0}(y'-x') \gamma^{\rho '} \GG_{0}(x'-z')
 \gamma^{\nu '} \GG_{0}(z'-y')]   \\  \nonumber 
 \times D_{\rho,\rho '}(x-x') D_{\mu,\mu '}(y-y') D_{\nu,\nu '}(z-z') 
 \left(\frac{ie}{2}(z'-y')^{\alpha}F_{\alpha\beta}(y'-x')^{\beta}\right)
 +(\, \mu, y \leftrightarrow \nu,z ),
\eea
where $D_{\alpha,\alpha '}(x-x')$ are the propagators of the photons. In Fourier 
space this is given by 
\bea
 \Gamma^{\rho\mu\nu}[x,y,z]\! \!&=& \! \!
 \int \frac{d^{4}p}{(2 \pi)^{4}} \frac{d^{4}p_{1}}{(2 \pi)^{4}} 
 \frac{d^{4}p_{2}}{(2 \pi)^{4}} 
 D_{\rho,\rho '}(p) D_{\mu,\mu '}(p_{1}) D_{\nu,\nu '}(p_{2}) 
 e^{ipx} e^{-ip_{1}y} e^{-ip_{2}z}                              \nonumber  \\
 && \times (2 \pi)^{4} \delta^{4}(p-p_{1}-p_{2})
 M^{\rho '\mu '\nu '}(p,p_{1},p_{2}).
\eea
Thus, the corresponding three-point amplitude reads 
\bea    \label{trianfase}
 M_{2}^{\rho\mu\nu}[p,p_{1},p_{2}]&=& \!\! -(i)^{3} e^{3} T \sum_{n}\int
 \frac{d^{3}k}{(2 \pi)^{3}} 
 \frac{ie}{2}\, F_{\alpha\beta}
 \frac{\partial}{\partial p_{1}^{\alpha}}\frac{\partial}{\partial p_{2}^{\beta}}
 \\ \nonumber 
 &&\Tr[
 \left( \gamma^{\mu}\GG_{0}(k+p_{1}) \gamma^{\rho} \GG_{0}(k-p_{2})
 \gamma^{\nu} \GG_{0}(k) \right)] 
 +(\, \mu,p_{1} \leftrightarrow \nu,p_{2}).  
\eea
Proceeding as we did before for the other contribution we obtain an identical 
result. Thus the total amplitudes are
\bea
 && M_{\parallel \to \perp \perp}=
 -\frac{e B_{y}}{m^{2}}\frac{e}{\omega} \: m_{t}^{2},
 \\
 && M_{\perp \to \parallel \, \perp}=
 \frac{e B_{y}}{m^{2}}\frac{e}{\omega_{1}} \: m_{t}^{2},
 \\
 && M_{\perp \to \perp \parallel}=
 \frac{e B_{y}}{m^{2}}\frac{e}{\omega_{2}} \: m_{t}^{2},
 \\
 && M_{\parallel \to \parallel \, \parallel}=
 \frac{e B_{y}}{m^{2}}\frac{e(\omega_{1}^{2}+\omega_{1}
 \omega_{2}+\omega_{2}^{2})}{\omega \omega_{1} \omega_{2}} \: m_{t}^{2},
 \\
 && M_{\parallel \to \parallel \, \perp}=M_{\parallel \to \perp \parallel}=
 M_{\perp \to \parallel \, \parallel}=M_{\perp \to \perp \perp}=0.
\eea
The amplitudes with an odd number of perpendicularly polarized photons vanish. 
These results can be written in a compact form
\be         \label{triangleB}
 M^{ijk}(p,p_{1},p_{2})=e^{2}\frac{m_{t}^{2}}{m^{2}} \left[
 \frac{1}{\omega_{2}}\: \delta^{ij}\: V^{k}
 -\frac{1}{\omega}\: \delta^{jk}\: V^{i}
 +\frac{1}{\omega_{1}}\: \delta^{ki}\: V^{j} \right],
\ee
where $V^{j}=\epsilon^{jkm}B_{k}\puni_{m}$ and $i,j,k$ can take the values $1,2$
for polarizations $\parallel,\perp$ respectively.

Now we turn to the contribution at order $B^{2}$ coming from the pentagon graph. 
The computation proceeds along the lines of the previous one. The contribution 
at second order in $B$ with no phases is computed as before. For the term linear 
in the phase we start from Eq.(\ref{trianfase}), but now the factor inside the 
trace must be replaced by its first order expression in $B$. Finally, for the 
second order contribution of the phase we have an expression similar to 
Eq.~(\ref{trianfase})
\bea    \label {trianfase2}
 M^{\rho\mu\nu}[p,p_{1},p_{2}]&=& \!\! -\frac{1}{2}(i)^{3} e^{3} T \sum_{n}\int
 \frac{d^{3}k}{(2 \pi)^{3}} 
 \left(\frac{ie}{2}\right)^{2}\, F_{\alpha\beta}F_{\sigma\eta}
 \frac{\partial}{\partial p_{1}^{\alpha}}\frac{\partial}{\partial p_{2}^{\beta}}
 \frac{\partial}{\partial p_{1}^{\sigma}}\frac{\partial}{\partial p_{2}^{\eta}}
 \\ \nonumber 
 &&\Tr[
 \left( \gamma^{\mu}\GG_{0}(k+p_{1}) \gamma^{\rho} \GG_{0}(k-p_{2})
 \gamma^{\nu} \GG_{0}(k) \right)] 
 +(\, \mu,p_{1} \leftrightarrow \nu,p_{2}).  
\eea
The results of each part alone are not very illustrative, so we give only the 
total amplitude
\bea
 && M_{\parallel \to \parallel \, \perp}=
 - i e^{3} \left(\frac{B_{y}B_{z}}{B_{cr}^{2}}\right) 
 \left(\frac{\omega_{1}\omega+3\omega_{2}^{2}}{\omega\omega_{1}\omega_{2}^{2}}\right) n_{e},
 \\
 && M_{\parallel \to \perp \parallel}=
 - i e^{3} \left(\frac{B_{y}B_{z}}{B_{cr}^{2}}\right) 
 \left(\frac{\omega_{2}\omega+3\omega_{1}^{2}}{\omega\omega_{1}^{2}\omega_{2}}\right) n_{e},
 \\
 && M_{\perp \to \parallel \, \parallel}=
  i e^{3} \left(\frac{B_{y}B_{z}}{B_{cr}^{2}}\right) 
 \left(\frac{3\omega_{1}^{2}+5\omega_{1}\omega_{2}+3\omega_{2}^{2}}
 {\omega^{2}\omega_{1}\omega_{2}}\right) n_{e},
 \\
 && M_{\perp \to \perp \perp}=
 - i e^{3} \left(\frac{B_{y}B_{z}}{B_{cr}^{2}}\right) 
 \left(\frac{\omega_{1}^{2}+\omega_{1}\omega_{2}+\omega_{2}^{2}}
 {\omega\omega_{1}\omega_{2}}\right)^{2} n_{e},
 \\
 && M_{\parallel \to \perp \perp}=M_{\perp \to \parallel \, \perp}=
 M_{\perp \to \perp \parallel}=M_{\parallel \to \parallel \, \parallel}=0,
\eea
with 
\be  \label{charden}
 n_{e}=2 \int \frac{d^{3}p}{(2 \pi)^{3}}
 (n_{F}(\varepsilon_{p}-\mu)-n_{F}(\varepsilon_{p}+\mu))
\ee
being the net electron density. The amplitudes for an even number of perpendicularly 
polarized photons vanish. These results can also be written in a compact form
\bea         \label{triangleB2}
 M^{ijk}(p,p_{1},p_{2})=&&-i e^{3}\: 
 \left(\frac{{\bf B}\cdot \pvuni}{B_{cr}^{2}}\right)\, n_{e} 
 \\ \nonumber
 &&\times \left[
  \left(\frac{1}{\omega^{2}}-\frac{3}{\omega_{1}\omega_{2}}\right) \delta^{jk}\: B_\perp^{i}
 +\left(\frac{1}{\omega_{1}^{2}}+\frac{3}{\omega\omega_{2}}\right) \delta^{ik}\: B_\perp^{j}
 +\left(\frac{1}{\omega_{2}^{2}}+\frac{3}{\omega\omega_{1}}\right) \delta^{ij}\: B_\perp^{k}
 \right],
\eea
where $B_\perp^{i}=B^{i}-({\bf B}\cdot \pvuni)\puni^{i}$ and $i,j,k$ can take 
the values $1,2$ for $\parallel,\perp$ respectively.

An important issue concerning the outlined computations is its consistency with 
gauge invariance. Following the same procedure, we have computed other components 
of the three-point amplitude and we have explicitely verified that the Ward 
identities in the collinear approximation for null vectors, $M^{0\mu\nu}-M^{3\mu\nu}=0$, 
are satisfied.


\section{The absorption coefficient}

In order to obtain the absorption coefficient a phase space integration with the 
proper measure has to be performed. When the effects of the magnetic field are 
considered in the dispersion relations, Adler~\cite{adler} showed by using arguments 
based on CP invariance and kinematics, that in the vacuum the only allowed channel 
is $\parallel \to \perp \, \perp$. As we have mentioned above, for high energy 
photons ($\omega \gg \omega_0$) propagating in a magnetized plasma, the influence 
of density or temperature on the dispersion relation is small and we can expect 
that the same channel is the only one allowed for contributions which are even 
in the chemical potential. For this decay we have
\be         \label{absorption}
 d\kappa_{\parallel \to \perp \perp}
 =\frac{1}{2}\frac{1}{2\omega 2 \omega_{1} 2 \omega_{2}}
 \frac{d^{3}k_{1}}{(2 \pi)^{3}}
 \frac{d^{3}k_{2}}{(2 \pi)^{3}}\; 
 (2 \pi)^{4}  \delta^{4}(k-k_{1}-k_{2})\times \vert M_{\parallel \to \perp \perp} \vert^{2},
\ee
where the enhancement Bose factors appearing in the probability have been replaced 
by one, because we consider the case $\omega_{1},\omega_{2} \gg T$. If dispersion 
due to the magnetic field is also neglected, the integrals reduce to 
$\int_{\omega_{0}}^{\omega-\omega_{0}} dk_{1}$, which can be approximated by the 
overall integral, so that 
\be   
 \kappa_{\parallel \to \perp \, \perp}=
 \frac{\vert M_{\parallel \to \perp \, \perp}\vert^{2}}{32 \pi \: \omega}=
 \frac{\alpha}{8} \left(\frac{B \sin \theta}{B_{cr}}\right)^{2}
 \left(\frac{m}{\omega}\right)^{3} \left(\frac{m_{t}}{m}\right)^{4} m.
\ee

For the pentagon contribution the kinematical constraints are the same as before, 
but the CP arguments now allow only procesess with an odd number of perpendi\-cu\-larly 
polarized photons. When both constraints are taken into account the only allowed 
channel is $\parallel \to \parallel_1 \perp_2$ or $\parallel \to \perp_1 \parallel_2$. 
The absorption coefficient is given by an expression similar to Eq.~(\ref{absorption}) 
with $\vert M_{\parallel \to \perp \perp} \vert^{2}$ replaced by 
$\vert M_{\parallel \to \parallel \perp} \vert^{2} + 
\vert M_{\parallel \to \perp \parallel} \vert^{2}$. 
Now, the integrand depends on $\omega_1$ and $\omega_2$ and hence the final result 
displays an explicit dependence on the cutoff
\be
 \kappa_{\parallel \to \parallel \, \perp}=
 \frac{\pi^{2}\alpha^{3}}{3} \left(\frac{B}{B_{cr}}\right)^{4} \sin^{2}(2\theta)
 \left(\frac{m}{\omega}\right)^{2} \left(\frac{m}{\omega_{0}}\right)^{3}
 \left(\frac{n_{e}}{m^{3}}\right)^{2} m.
\ee
We stress again that an application of the above formulae for low energy photons 
is misleading, as the formulae themselves show. At low energy, the propagating 
modes in the plasma are longitudinal or transverse plasmons whose dispersion 
relations differ significantly from that we have assumed, $\omega=p$. 

We show the explicit dependence on the plasma parameters for two opposite regimes. 
For a completely degenerated electron plasma we have 
\bea  \label{degewidth}
 \kappa_{\parallel \to \perp \, \perp}&=&
 \frac{\alpha^{3}}{2\pi^{2}}\left(\frac{B\sin \theta}{B_{cr}}\right)^{2}
 \left( \frac{m}{\omega} \right)^{3}
 \times \left[\ln\left(\frac{\mu+\sqrt{\mu^{2}-m^{2}}}{m}\right)-
 \frac{\mu}{m}\sqrt{\frac{\mu^{2}-m^{2}}{m^{2}}} \right]^{2} m, \\ 
 \kappa_{\parallel \to \parallel \, \perp}&=&
 \frac{\alpha^{3}}{27\pi^{2}}\left(\frac{B}{B_{cr}}\right)^{4} \sin^{2}(2\theta)
 \left(\frac{m}{\omega}\right)^{2} \left(\frac{m}{\omega_{0}}\right)^{3}
 \left(\frac{\mu^{2}-m^{2}}{m^{2}}\right)^{3} m,
\eea
and for high temperature $T \gg m$ and $\mu=0$, we obtain at order $B^{2}$
\be
 \kappa_{\parallel \to \perp \, \perp}=
 \frac{\pi^{2}\alpha^{3}}{18}\, \left(\frac{B\sin \theta}{B_{cr}}\right)^{2}
 \left( \frac{T}{\omega} \right)^{3}\, T.
\ee
Obviously, there is no contribution at order $B^{4}$.

In order to find an estimate of matter effects in photon splitting, we compute 
the absorption coefficient for two astrophysical objects. For a white dwarf we 
consider the following values of the parameters~\cite{shapiro} 
\be
 \frac{B}{B_{cr}}\sim 2 \cdot 10^{-6}, \ \ \ \frac{T}{m}\sim 2 \cdot 10^{-3},  \ \ \ 
 \frac{\mu}{m}\sim 1.02. 
\ee
The transverse mass and the charge density can be estimated with formulas (\ref{mt},
\ref{charden}), giving $m_{t} \sim 5 \cdot 10^{-3}\, m,\ n_{e}\sim 2.7\cdot10^{-4} m^3$. 
In order to satisfy (\ref{masst}) we need $\omega \gg 2.5\cdot 10^{5}\, m$. 
Taking $\theta =\pi/4$ and $\omega \sim 10^{6}\, m$ we get the following estimates
\be
 \kappa_{\parallel \to \perp \, \perp} \simeq 10^{-42} \, m, \ \ \  
 \kappa_{\parallel \to \parallel \, \perp} \simeq 10^{-64} \, m.
\ee
The value in the vacuum from the hexagon is
$\kappa_{\parallel \to \perp \, \perp} \simeq 3.5 \cdot 10^{-17} \, m$.
Here, the effects of matter are completely irrelevant. For a neutron star we can 
take the values~\cite{shapiro} 
\be
 \frac{B}{B_{cr}}\sim 2 \cdot 10^{-1}, \ \ \ \frac{T}{m}\sim 1,   \ \ \ 
 \frac{\mu}{m}\sim 600.
\ee
Now we have $m_{t} \sim 40.9\, m$, $n_{e} \sim 7.3\cdot10^{6}\, m^{3}$ and 
$\omega_{0} \sim 2\cdot 10^{4}\, m$. Again, taking $\theta=\pi/4$ and 
$\omega \sim 10^{5}\, m$ we get the estimate
\be
 \kappa_{\parallel \to \perp \, \perp} \simeq 5 \cdot 10^{-14} \, m, \ \ \ 
 \kappa_{\parallel \to \parallel \, \perp} \simeq 1.4 \cdot 10^{-18} \, m, 
\ee
while the value of the hexagon in the vacuum is
$\kappa_{\parallel \to \perp \, \perp} \simeq 3.5 \cdot 10^{8} \, m$. 
We see that the effects of the medium are neglible by many orders of magnitude.
In addition, the results for the neutron star do not change appreciably in the
range of temperatures from $T/m=10^{-3}$ to $T/m=1$.


\section{Summary}

In this work we have computed the contribution of order $B^{2}$ and $B^{4}$ to 
the photon splitting process at finite temperature or density in the collinear
aproximation. We have computed directly the three-photon amplitude using fermionic 
propagators in the presence of a magnetic external field for photons with 
$\omega=p$. The phase of the propagator contributes non trivially to the amplitude. 
In order to have photons with $\omega \simeq p$ we have restricted ourselves to 
photons with $\omega \gg \omega_{0}$ so the effects of the medium on the dispersion 
relations can be neglected compared to those of the magnetic field. This condition 
leads to consider very high energy photons, for which the effects of the medium 
are neglible compared with the perturbative vacuum contribution coming from the 
hexagon, as seen in explicit computations. It is not excluded that for lower 
energy photons, when the full effects of the dispersion relations must be taken 
into account, the box contribution is more significant. 

A natural extension of this work is the study of the matter-induced three-point 
vertex at strong field, a regime with possible astrophysical applications. At
strong field, the modification of the dispersion relations due to vacuum effects
play a more significant role~\cite{chistyakov} than at weak fields.


\section*{Acknowledgments}

We would like to thank M. A. Go\~ni for useful discussions. 
This work is partially supported by UPV/EHU grant 063.310-EB187/98 and 
CICYT AEN 99-0315. 
J. M. Mart\'\i nez Resco is supported by a Basque Government grant.



\newpage


\begin{figure}
\centering
\leavevmode
\epsfysize=4cm
\epsfbox{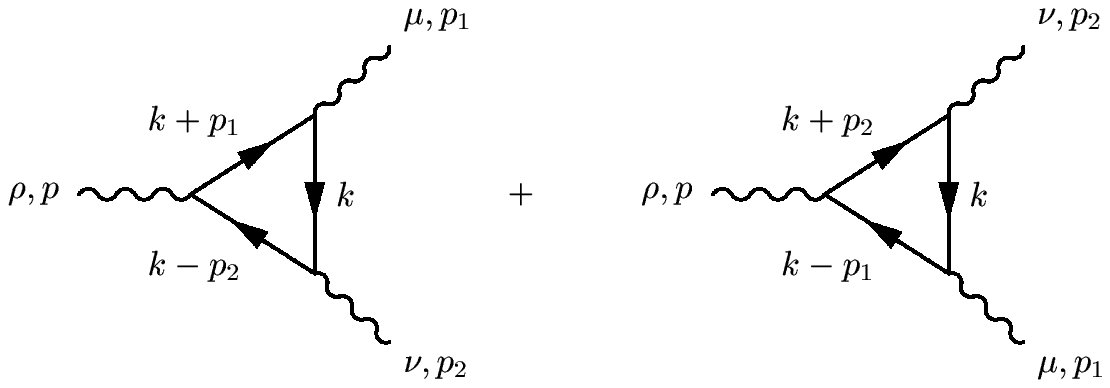}
\caption{The photon splitting amplitude. The solid lines represents fermionic 
propagators in the presence of a magnetic external field.}
\label{fgtri}
\end{figure}


\end{document}